# Effect of Edge Roughness on Electronic Transport in Graphene Nanoribbon Channel Metal Oxide Semiconductor Field-Effect Transistors

D. Basu[1], M. J. Gilbert[1], L. F. Register[1], A. H. MacDonald[2] and S. K. Banerjee[1]

*1) Microelectronics Research Center, The University of Texas at Austin, 10100 Burnet Road, Bldg. 160, MER 1.606B/R9900, Austin, TX 78758-4445*

*2) Department of Physics, The University of Texas at Austin, 1 University Station C1600, Austin, TX 78712-0264*

## Abstract:

Results of quantum mechanical simulations of the influence of edge disorder on transport in graphene nanoribbon metal oxide semiconductor field-effect transistors (MOSFETs) are reported. The addition of edge disorder significantly reduces ON-state currents and increases OFF-state currents, and introduces wide variability across devices. These effects decrease as ribbon widths increase and as edges become smoother. However the bandgap decreases with increasing width, thereby increasing the band-to-band tunneling mediated subthreshold leakage current even with perfect nanoribbons. These results suggest that without atomically precise edge control during fabrication, MOSFET performance gains through use of graphene will be difficult to achieve.

Graphene has recently generated considerable interest as a semiconductor that can potentially be used to help advance the silicon technology roadmap.[1] Although bulk graphene sheets are metallic, nanoscale graphene ribbons have bandgaps. The potential for performance gains in MOSFETs employing semiconducting graphene nanoribbon channels have been studied theoretically using both effective-mass models[2] and more involved full band models.[3-5] The performance of graphene channel with various constrictions has also been investigated.[6] Initial results from experiments have also demonstrated that it is indeed possible to fabricate field effect devices using graphene as the channel material in a conventional complementary MOS (CMOS) process flow.[7] Experimentalists have been able to pattern graphene into ribbons with widths of the order of tens of nm.[8] Theoretical studies of graphene nanoribbons as a channel material for MOSFETs have focused mainly on perfect armchair graphene ribbons with widths as small as 1.4 nm.[4] Ribbons this narrow have energy gaps that are sufficiently large for use as a semiconducting channel material. In practice, though, it is difficult to fabricate samples of graphene with perfect edges,[9] and one can expect the effect of non-idealities to increase as the width is scaled down. In this work, we report on the effect of these edge imperfections in graphene nanoribbon MOSFETs as observed through full band ballistic quantum transport simulation. A theoretical analysis of the role of edge disorder in transport in simple conduction channels of graphene ribbons with a different emphasis appeared as this work was in progress.[10]

The carbon (C) to carbon in-plane bond in graphene is strong, making graphene sheets quite robust. Edges, however, tend to be rough.[8,9] In particular, carbon atom vacancies with respect to what would be a perfect armchair edge appear difficult to avoid. Edge vacancies give rise to steps along the edge.[9] In this work we model the edges by defining a correlation number $r$ as the fraction of times the state of an edge site is identical to the corresponding site in the preceding slice, where we consider the two dimensional (2D) graphene nanoribbon as a series of single atom thick slices along the transport direction. Thus, with $r = 0.9$, steps in the edge occur 10% of the time, either from a series of C atoms to vacant sites or vice versa. With $r = 0.5$, edge sites are randomly vacant. With $r = 1.0$, a perfect armchair edge is achieved. A graphene plane 10.5 nm long and 4.18 nm wide with $r = 0.9$ is shown in Fig. 1(a).

We represent the graphene layer by its $\pi$ orbital nearest neighbor tight binding model

$$H_{ij} = tN_{ij} - q\phi_i\delta_{ij}. \qquad (1)$$

Here $N_{ij}$ is 1 for the honeycomb lattice nearest neighbors, and is zero otherwise, and $t$ is the nearest neighbor C-C hopping energy and is –3.03 eV.[11] $\phi$ is the self-consistent electrostatic potential obtained by solving Poisson's equation in three dimensions (3D), and $q$ is the electronic charge. We inject eigenmodes from the matched leads at the source and drain contacts, and use recursive scattering matrices to propagate the wavefunctions through the device from source (drain) to drain (source) in real space. Current is calculated by integrating the transmission coefficients over energy with a Fermi function weight. The details of our full band quantum transport method, which follows Ref. 12 closely, will be described elsewhere.[13] Following Ref. 4, we consider a dual gate MOSFET structure where conduction occurs in the graphene layer sandwiched between top and bottom silicon oxide ($SiO_2$) layers. Source and drain regions are n+ doped to a concentration of $10^{13}$ cm$^{-2}$, and are contacted by perfectly matched semi-infinite graphene leads. (Consideration of Schottky contacts is a separate issue beyond the scope of this work.) Fig. 1(b) shows a schematic of the graphene nanoribbon dual gate MOSFETs that we study. The channel region is nominally undoped.

The total "transmission" ($T$)— here defined as the sum over the individual transmission probabilities for each lead subband— under flatband conditions ($\phi \equiv 0$) as a function of total energy ($E$) are shown in Fig. 2 for imperfect nanoribbon channels of varying nominal widths ($W_{ch}$). In each case, atomically identical edges with edge roughness characterized by $r = 0.90$ were used. Results for perfect edges ($r = 1$) are shown for comparison. (Variations among nanoribbons with the same roughness parameter $r$ but atomically different edges are addressed subsequently.) The disorder introduces scattering. As the width is scaled down the impact of the edge disorder increases and transmission goes down. The result is very poor transmission in the narrow ribbons. $T(E)$ results (not shown in Fig. 2.) for $W_{ch}$ =1.72 nm, which is still wider than the $W_{ch}$ value considered in Ref. 4, are essentially zero up to at least $E = 0.5$ eV. Fig. 3 shows transmission characteristics for 7.63 nm wide graphene channels with five different values of the edge roughness parameter $r$. Transmission falls drastically as the correlation falls below 0.99 and the number of steps along the edge increases.

The drain current per unit width, $I_D/W_{ch}$, vs. gate voltage ($V_G$) relations for dual gate MOSFETs with graphene nanoribbon channels with atomically identical edges characterized by $r = 0.9$ (we use the same edge configuration here as for the simulations of Fig. 2) but different widths are shown in Fig. 4. Notably, in these electrostatically self-consistent simulations, accumulation of charge on the C atoms in the vicinity of the steps, results in variations in the potential $\phi$ and, thus, further scattering. As can be seen in Fig. 4(a), the degradation of drain current above threshold is large, although, as expected, relatively smallest for the $W_{ch} = 15.74$ nm device. However, as seen in Fig. 4(b), only for the narrowest ribbon width, $W_{ch} = 4.18$ nm, is the subthreshold behavior marginally acceptable for a MOSFET even with atomically smooth nanoribbon edges, a consequence of band-to-band tunneling mediated leakage currents. And, also as seen in Fig. 4(b), edge roughness can increase leakage in the nominally OFF state. While the graphene nanoribbons considered here are nominally semiconducting, that is they are so in the limit of no edge roughness, the band gap of graphene is a sensitive function of the channel width. For armchair nanoribbons of constant width, in the nearest neighbor tight-binding model that is used, one obtains a metallic energy dispersion relation if the number of C atoms along the width is $3l+1$ where $l$ is an integer.[14] With only some of the edge sites vacant, one gets an admixture of different widths and the edge roughness results in quasi-localized defect states within the band gap.

Edge roughness also leads to significant variability in performance among devices with different atomic edge configurations even with the same correlation parameter $r$ and channel width. This variability can be seen in Fig. 5 in $I_D/W_{ch}$ for two set of devices with $W_{ch} = 4.18$ nm, one set with $r = 0.99$ and one set with $r = 0.5$ values. The error bars represent plus or minus the standard deviation— a range less than that that would have to be allowed for in circuit design— for a set of ten different random edge configurations with identical $r$. Even with $r = 0.99$ the variability is great.

These results associated with edge disorder, greatly reduced ON-state currents and increased OFF-state leakage currents combined with wide variability among devices with the same degree of edge roughness ($r$) but atomically different edge configurations, suggest a need for fabrication of graphene nanoribbon channels with atomically near perfect edges to a realize the potential for performance gains in MOSFETs previously reported.[2-5] And the use of wide channel devices to minimize edge effects is not an alternative, as even devices with perfect nanoribbon edges, exhibit prohibitive off-state leakage currents.

This work was supported in part by the NRI SWAN Center, DARPA and the GRC.

## References:

[1] International Technology roadmap for semiconductors, 2005 update. Available from http://www.itrs.net/

[2] G.-C. Liang, N. Neophytos, D. Nikonov, and M. Lundstrom, IEEE Trans. Electron Devices 54, 677 (2007).

[3] G. Fiori and G. Iannaccone, IEEE Electron Device Lett. 28, 760 (2007).

[4] G.-C. Liang, N. Neophytos, D. Nikonov, and M. Lundstrom, J. Appl. Phys. 102,054307 (2007).

[5] Y. Ouyang, Y. Yoon, and J. Guo, IEEE Trans. Electron Devices 54, 2223 (2007).

[6] F. Munoz-Rojas, J. Fernandez-Rossier, L. Brey and J. J. Palacios, arXiv:0707.0375v1 [cond-mat.mes-hall]

[7] M. C. Lemme, T. J. Echtermeyer, M. Baus, and H. Kurz, IEEE Electron Device Lett. 28, 282 (2007).

[8] M. Y. Han, B. özyilmaz, Y. Zhang, and P. Kim, Phys. Rev. Lett., 206805 (2007).

[9] C. Berger, Z. Song, X. Li, X. Wu, N. Brown, C. Naud, D. Mayou, T. Li, J. Hass, A. N. Marchenkov, E. H. Conrad, P. N. First, and W. A. de Heer, Science 312, 1191 (2006).

[10] I. Martin and Y. M. Blanter, arXiv:0705.0532v2 [cond-mat.mes-hall]

[11] R. Saito, G. Dresselhaus, and M. Dresselhaus, *Physical Properties of Carbon Nanotubes*, Imperial College Press, London, (1998).

[12] T. Usuki, M. Saito, M. Takatsu, R. A. Kiehl, and N. Yokoyama, Phys. Rev. B 52, 8244 (1995).

[13] D. Basu, M. J. Gilbert, L. F. Register and S. K. Banerjee, *An Efficient Method for Quantum Transport Calculations in Nanostructures using Full Band Structure*, to be submitted to J. Appl. Phys.

[14] More realistic and rigorous ab-initio calculations however indicate that all graphene nanoribbons with perfect armchair edges have energy gaps. See Y-W. Son, M. L. Cohen, and S. G. Louie, Phys. Rev. Lett., 216803 (2007).

**Figure Captions:**

**Fig. 1.** (a) Armchair graphene nanoribbon used as the channel material in a MOSFET, showing vacant sites along the edges where C atoms (black dots) are missing. (b) Schematic of the simulated device structure (side view). For clarity, the nanoribbon of (a) is shorter (10.5 nm) than that used in the simulated MOSFET device of (b).

**Fig. 2.** Transmission $T(E)$ as a function of incident energy $E$ across graphene channels having identically rough edges. Steps show perfect transmission for ideal armchair edges of corresponding width.

**Fig. 3.** $T(E)$ as a function of $E$ for a 7.63 nm wide graphene channel having different roughness at the edges. $r = 0.5$ has edge sites randomly vacant and $r = 1.0$ has a perfect armchair edge.

**Fig. 4.** $I_D$–$V_G$ characteristics on (a) linear-linear and (b) log-linear scale for the MOSFET structure of Fig. 1, for three different channel widths, showing performance degradation for channels with rough edges (dashed lines, solid symbols) from the ideal ballistic devices (solid lines, open symbols). Drain voltage $V_D$ for all of these simulations is 0.2 V.

**Fig. 5.** $I_D$–$V_G$ for the dual gate MOSFET with $W_{ch} = 4.18$ nm, and using different values of edge roughness parameter $r$. Error bars plotted indicate standard deviation in $I_D$ across ten randomly different edges having macroscopically same values of $r$. $V_D = 0.3$ V for all of these simulations.

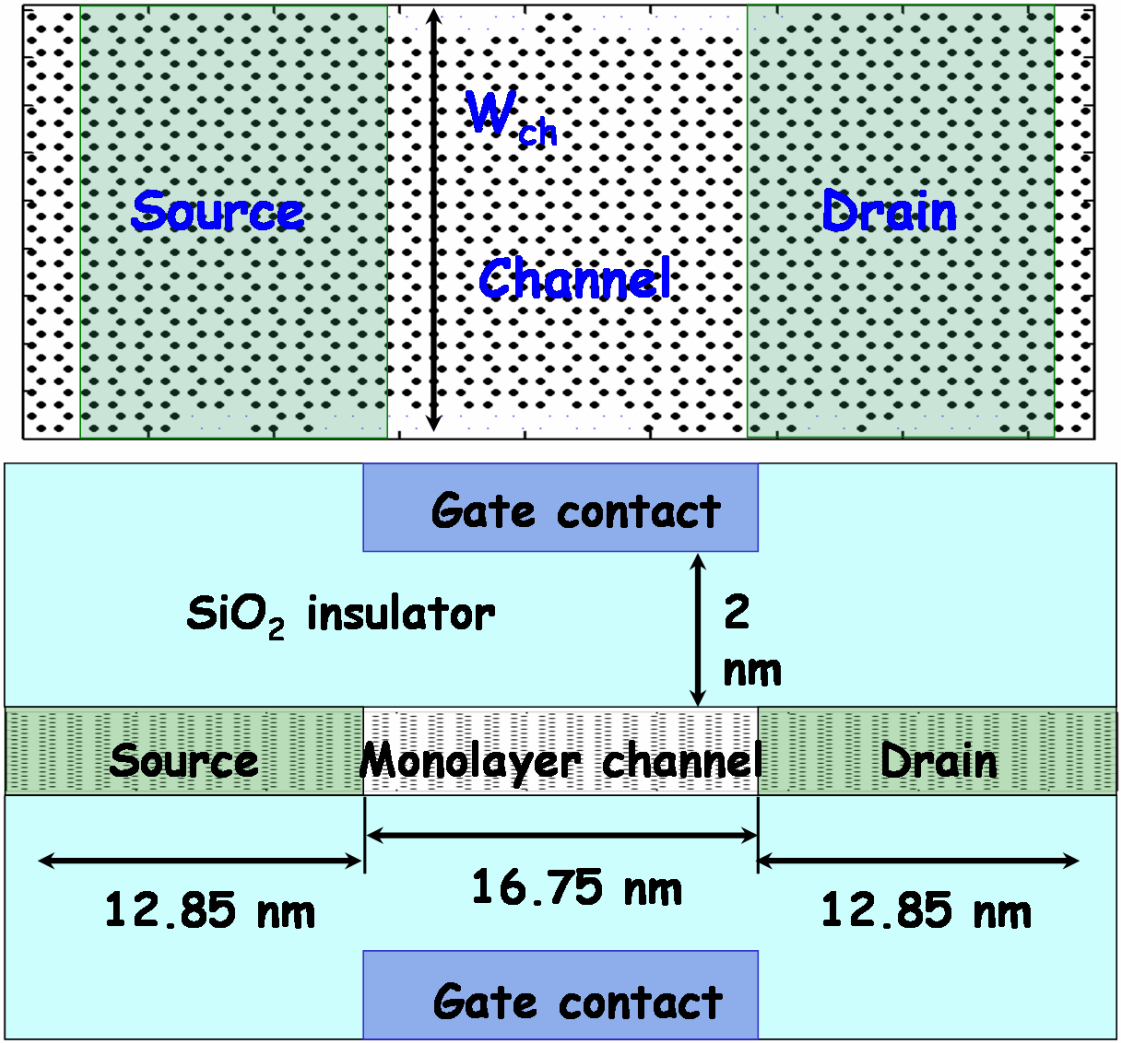

Fig. 1.

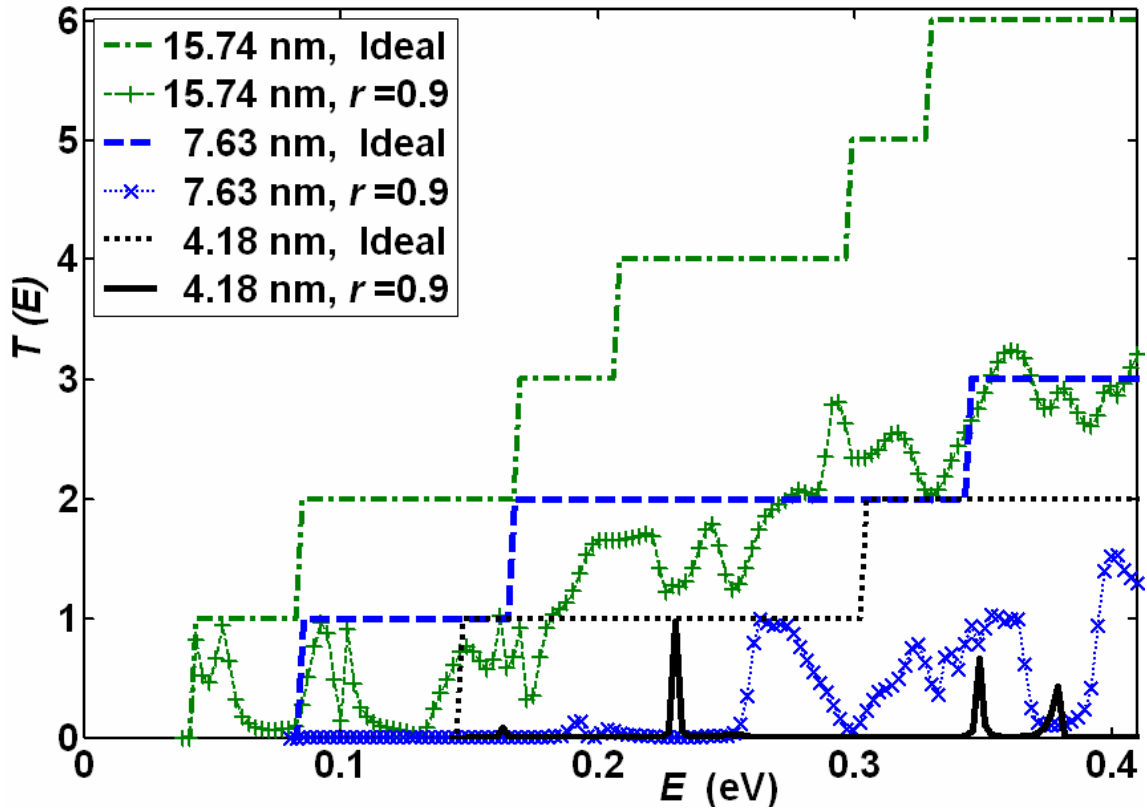

**Fig. 2.**

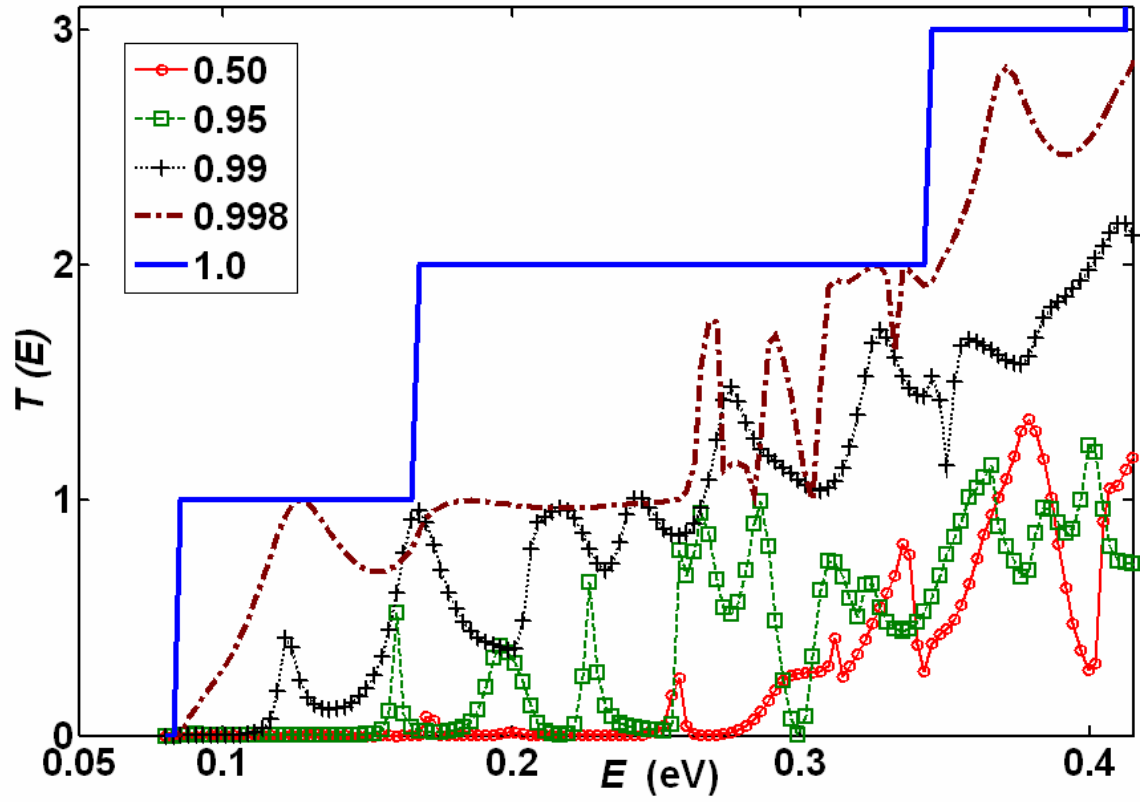

Fig. 3.

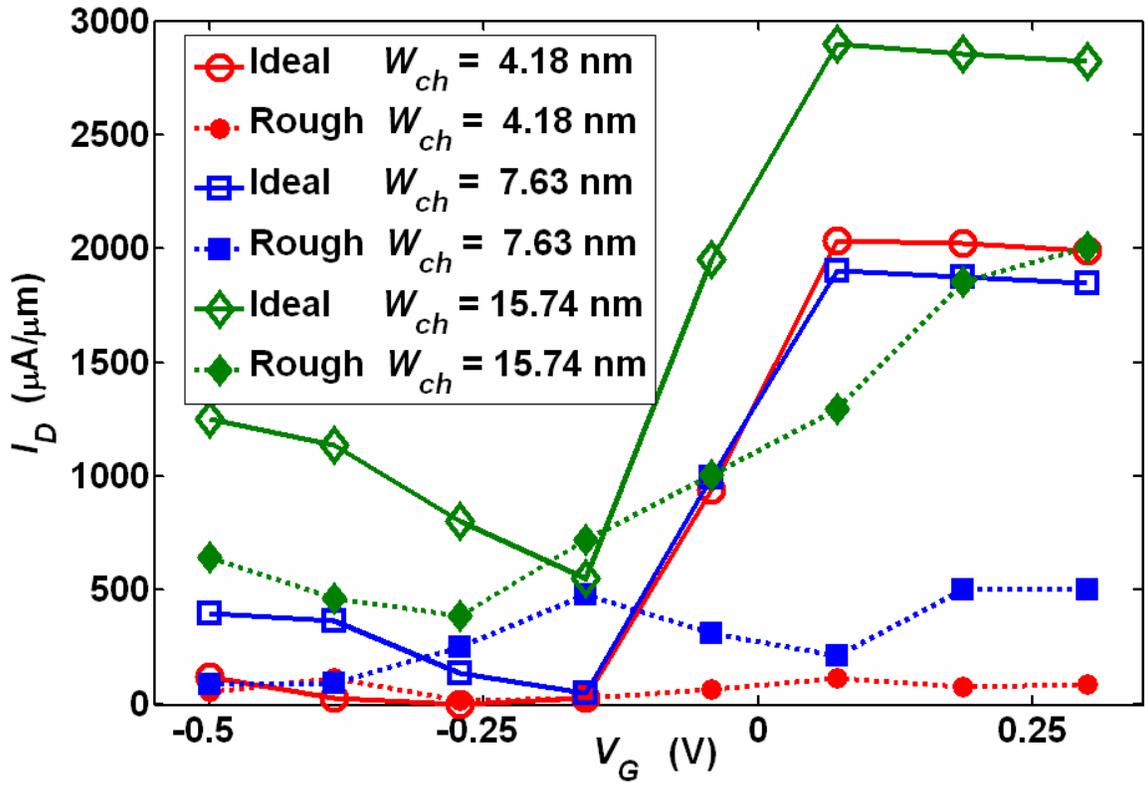

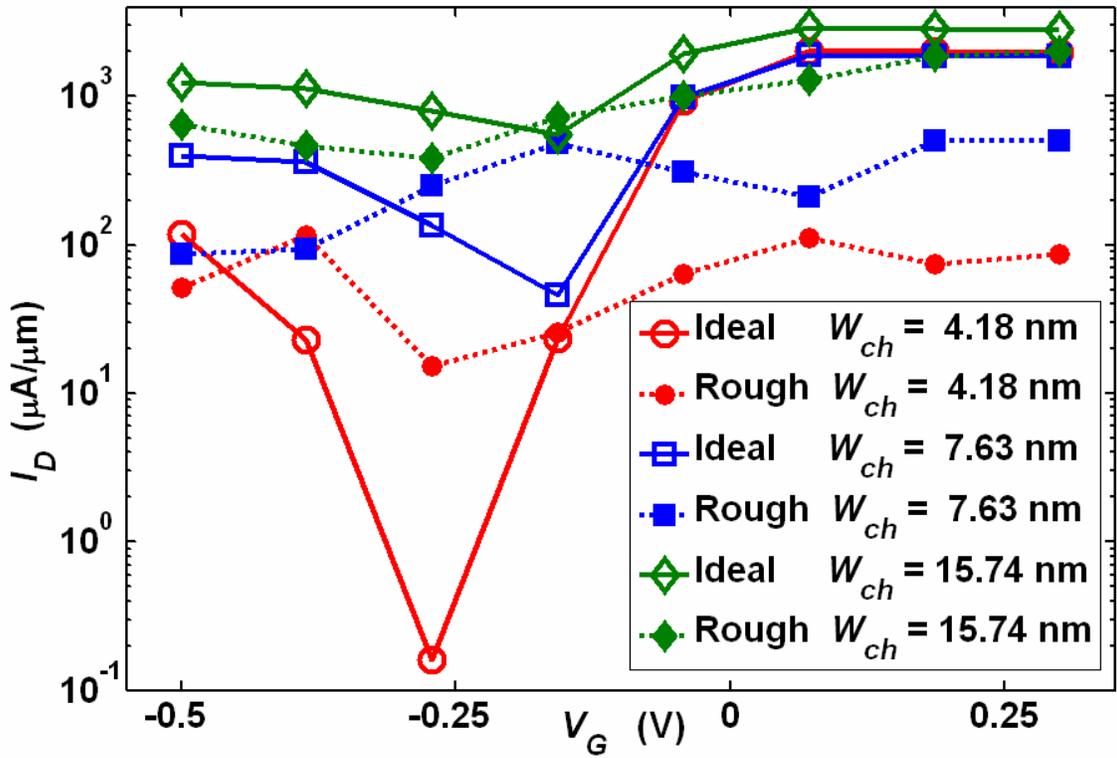

Fig. 4.

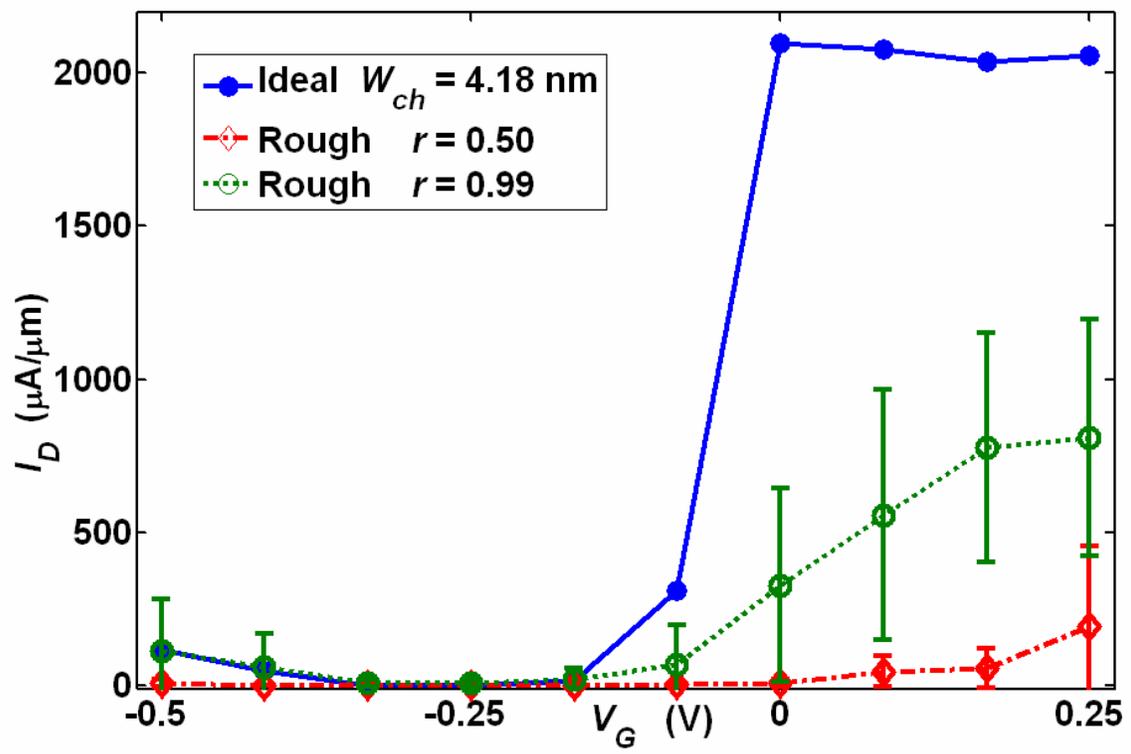

Fig. 5.